\documentstyle[12pt,titlepage]{article} 
\title{ Heavy to Light Semileptonic Transitions in the
Heavy Quark Effective Theory}
\author{Gustavo Burdman}          
\date{October, 1992}   
\def\_{\rule{.3em}{.15ex}}  

\begin{document}
\begin{titlepage}
 \begin{flushright}
  UMHEP-382
 \end{flushright}
 \begin{center}
  \vspace{0.75in}
  {\bf {\LARGE Heavy to Light Semileptonic Transitions in the
Heavy Quark Effective Theory} \\
  \vspace{0.75in}
  Gustavo Burdman} \\
  Department of Physics and Astronomy \\
  University of Massachusetts \\
  Amherst, MA 01003 \\
  \vspace{1in}
  ABSTRACT \\
  \vspace{0.5in}
  \end{center}
  \begin{quotation}
  \noindent The scaling behavior of semileptonic form-factors
  in Heavy to Light transitions is studied in the Heavy Quark
Effective Theory. In the case of $H\rightarrow \pi e\nu$ it is
shown that the same scaling violations affecting the heavy meson decay
constant will be present in the semileptonic form-factors.

  \end{quotation}
\end{titlepage}


\newcommand{\da}{\mbox{$\scriptscriptstyle \dag$}}
\newcommand{\lag}{\mbox{$\cal L$}}
\newcommand{\tr}{\mbox{\rm Tr\space}}
\newcommand{\fc}{\mbox{${\widetilde F}_\pi ^2$}}
\newcommand{\ns}{\textstyle}
\newcommand{\si}{\scriptstyle}

\section {Introduction}

The Heavy Quark Limit (HQL), in which the mass of the heavy quark
$m_Q$ is much larger than the typical scale of hadronic
interactions, has been the object of intense study \cite{HQL}. The
symmetries arising in this limit, $SU(2)_{flavor}\times SU(2)_{spin}$,
have been successfully applied to the matrix elements involved in
transitions between heavy mesons, reducing the number of
form-factors to one universal function that contains all the long
distance effects. In the Heavy-to-Light transitions, although the
number of form-factors cannot be reduced by symmetry arguments,
the HQL can be used to relate $D\rightarrow L$ to $B\rightarrow L$
decays, where $L=\pi, \rho, \omega,\ldots$.
This is potentially useful to extract the CKM matrix element
$|V_{ub}|$ by using $D$ decay data plus the well defined scaling
behavior of the form-factors with the heavy mass in the HQL \cite{IW}.
Several issues should be addressed in the implementation of this
program. First, the $D$ decay data will not cover the full kinematic
range in the corresponding $B$ decay, as a consequence of which a
large portion of the $B$ data will be excluded. Second, the presence
of poles near the physical region, particularly in the $\pi$ modes,
introduces a different scaling behavior and, because they enter in $D$
and $B$ decays with different weights, they should be taken into
account. However, the most pressing issue concerning the extraction
of $|V_{ub}|$ by using HQL arguments is the size of the corrections
to the symmetry limit. In general, $1/m_Q$ corections will result
in scaling violations in the semileptonic form-factors. These
corrections are thought to be important in heavy meson decay
constants, a result suggested by lattice calculations \cite{lat1}.
It is then imperative to
confront this issue in semileptonic form-factors for
$H\rightarrow L$ transitions, for the existence of potentially
large scaling violations could introduce an uncertainty in the
scaling procedure that would seriously undermine our ability
to extract $|V_{ub}|$ within the framework of a controlled
approximation.

The Heavy Quark Effective Theory (HQET) \cite{HQET} is a convenient
framework to systematically take into account
the $1/m_Q$ corrections \cite{Mcor,Golden,Neu}.
In this letter we will show the structure of the $1/m_Q$
corrections to semileptonic form-factors in $H\rightarrow L$
transitions. We will focus on the case $L=\pi$, where the
formulation of its chiral couplings to heavy mesons \cite{Wise,BD}
gives a relation that can be used
to estimate the size of the scaling violation effects in terms of the
heavy quark symmetry breaking in the heavy meson decay constant.

\section {Semileptonic form-factors in the HQET}

In the HQL, form-factors entering in the semileptonic decays of
heavy mesons to light mesons have a definite scaling with the
heavy mass. The matrix element for the decay to a pion can be
written as

\begin{equation}
<\pi({\bf p})|\bar{q}\gamma_{\mu} Q|M({\bf P})> =
f_+(q^2) (P+p)_{\mu} +f_-(q^2) (P-p)_{\mu} \label{def}
\end{equation}

\noindent with the normalization of states given by

\begin{equation}
 <M({\bf P})|M({\bf P'})> = 2m_M \delta^{3}({\bf P}-
{\bf P'}) \label{nor}
\end{equation}

\noindent In the HQL the left hand-side of (\ref{def}) has to be
independent
of the heavy mass, after properly dividing by the factor $\sqrt{2m_M}$
from the normalization of the heavy meson. This automatically
implies \cite{IW}

\begin{eqnarray}
f_+ + f_- & \sim & m^{-1/2}_{M} \nonumber \\
          &      &         \label{sca1}   \\
f_+ - f_- & \sim & m^{1/2}_{M}  \nonumber
\end{eqnarray}

\noindent This simple scaling behavior is, however, affected by
corrections
that are multiplied by inverse powers of the heavy mass (as well as by
QCD corrections). The HQET \cite{HQET} provides a systematic approach
to these
corrections. The heavy quark field in the effective theory is defined,
in terms of the quark field in the full theory $Q(x)$, as

\begin{equation}
h_Q(v,x)=\exp^{im_Q\not v v.x} Q(x) \label{eff1}
\end{equation}

\noindent with $v$ the velocity of the heavy quark. In this way,
derivatives
on $h_Q(x)$ are `small' or independent of the heavy mass

\begin{equation}
i\!\!\not \partial h_Q(x)=(\not P - m_Q\!\!\not v)h_Q(v,x)=\not k h_Q(v,x)
 \label{eff2}
\end{equation}

\noindent
where $k$ is the off-shellness of the heavy quark, and it is of the
order of the $\Lambda_{QCD}$ scale for soft processes. In principle,
perturbative hard gluon exchange could upset this description in terms of
velocities. However, in practice, perturbative off-shell effects seem
small throughout the kinematic range relevant for $D$ and $B$ decays
\cite{Isgur,BD1}.

The effective Lagrangian is then written in terms of $h_Q$ as an
expansion in $1/m_Q$. Neglecting terms involving operators of
dimension six or higher \cite{Mcor}

\begin{equation}
{\cal L}_{eff}=\bar{h}_Q i v.D h_Q  +
\frac{1}{2m_Q}\bar{h}_Q [D^2 + \frac{1}{2}g_s \sigma^{\mu \nu}
G_{\mu \nu}] h_Q    \label{eff3}
\end{equation}

\noindent
where $D_{\mu}=\partial_{\mu} - ig_s A_{\mu}$
is the usual covariant derivative and
$G_{\mu \nu}=[D_{\mu},D_{\nu}]/ig_s$. The last two terms in (\ref{eff3})
break the $SU(2)_{flavor}$ symmetry due to the explicit presence of $m_Q$,
whereas only the last term breaks $SU(2)_{spin}$. Additional $m_Q$
dependence will appear when QCD corrections are taken into account.

The current operators will also have an expansion in $1/m_Q$. For the
Heavy-Light current, at tree level, we have \cite{Mcor}

\begin{equation}
\bar{q}\Gamma Q = \bar{q}\Gamma h_Q + \frac{1}{2m_Q}
\bar{q}\Gamma i\!\!\!\not D h_Q    \label{cur}
\end{equation}

\noindent where contributions from operators of dimension five or
higher are neglected.

To obtain the form-factors in the HQET we need the matrix elements
of the operators in the right-hand side of (\ref{cur}). The leading
contribution corresponds to the operator $Q_0 = \bar{q} \Gamma h_Q$.
Using the trace formalism \cite{Bj} this can be written as

\begin{equation}
<\pi({\bf p})|\bar{q}\Gamma h_Q|M({\bf v})> =
Tr [\gamma_5(A\!\not v + B\!\not p) \Gamma {\cal M}(v)]  \label{mat1}
\end{equation}

\noindent Here the heavy meson with velocity $v$ is represented by

\begin{equation}
{\cal M}(v) = \sqrt{m_M} \, \frac{(1+\not v)}{2} \, \gamma_5  \label{MWF}
\end{equation}

\noindent
It satisfies the conditions $\frac{(1+\not v)}{2}{\cal M}(v)=
{\cal M}(v)$ and $\frac{(1-\not v)}{2}{\cal M}(v)=0$. The explicit
mass dependence in (\ref{MWF}) is in correspondence with the
normalization (\ref{nor}). The form-factors $A$ and $B$ are
independent of $m_M$. They are functions of $v.p$, the light hadron
energy in the heavy meson rest frame. The expression (\ref{mat1}) is
completely general and independent of the Dirac structure. In particular,
for the vector current it will give the form-factors defined in
(\ref{def}) to leading order in the HQET.
The naive scaling
of (\ref{sca1}) is then recovered in the form:

\begin{eqnarray}
f_+ + f_- & = & 2A\, m_{M}^{-1/2}  \nonumber \\
          &   &          \label{sca2} \\
f_+ - f_- & = & 2B\, m_{M}^{1/2}   \nonumber
\end{eqnarray}

At tree level, the $1/m_Q$ corrections to (\ref{sca2})
come exclusively from the second term in (\ref{cur}), the operator
$Q_1 = \bar{q}\Gamma i\!\!\!\not D h_Q$. Dimension four operators
can be written in a general form by absorbing the Dirac matrix
contracting the covariant derivative into a general Dirac structure
$\Gamma$. Their matrix elements can then be expressed as

\begin{equation}
<\pi({\bf p})|\bar{q}\Gamma iD_{\alpha}h_Q|M({\bf v})> =
Tr[(F_1v_{\alpha} + F_2\gamma_{\alpha} + F_3 p_{\alpha})\gamma_5
\Gamma {\cal M}(v)] \label{mat2}
\end{equation}

\noindent
where the $F_i$'s are new long distance parameters, functions of
$v.p$ but independent of the heavy mass. In what follows we will
apply the methods that are used in Ref.\cite{Neu} for the
case of the decay constant in order to eliminate these new parameters
in favor of $A$ and $B$.

\noindent
First, the equation of motion for the heavy quark, to leading order, is
$iv.D h_Q = 0$. When this is applied to (\ref{mat2}) we obtain the
constraint

\begin{equation}
F_1 -  F_2 + v.p F_3 = 0 \label{con1}
\end{equation}

\noindent
The corrections to (\ref{con1}) are not relevant to the order we are
working.

\noindent
Furthermore, the $F_i$'s can be related
to the leading form factors $A$ and $B$ by making use of the fact
that both equationsns (\ref{mat1}) and (\ref{mat2}) are independent
of the Dirac structure of the currents. If we write

\begin{equation}
\bar{q}\Gamma iD_{\alpha} h_Q = i\partial_{\alpha}(\bar{q}\Gamma h_Q)
- (-iD_{\alpha})\bar{q} \Gamma h_Q  \label{parts}
\end{equation}

\noindent
and then set $\Gamma=\gamma^{\alpha} \hat{\Gamma}$, the last term in
(\ref{parts}) will vanish in the limit $m_q=0$ due to the equation
of motion for the light quark. In this way we can write

\begin{eqnarray}
<\pi({\bf p})|\bar{q}\gamma^{\alpha}\hat{\Gamma}
iD_{\alpha}h_ Q|M({\bf v})> & = &
i\partial_{\alpha}<\pi({\bf p})|\bar{q}\gamma^{\alpha}\hat{\Gamma}
h_Q|M({\bf v})> \\
   & = & \overline{\Lambda}v_{\alpha}
<\pi({\bf p})|\bar{q}\gamma^{\alpha}\hat{\Gamma}
h_Q|M({\bf v})>  \label{rel1}
\end{eqnarray}

\noindent
Here $\overline{\Lambda}=m_M - m_Q$ is a low energy parameter that
expresses the fact that in the effective theory the current only
carries a small momentum.

Using (\ref{mat1}), (\ref{mat2}) and (\ref{rel1}), we obtain two
equations relating the $F_i$'s with the leading form-factors.
Together with (\ref{con1}) they give

\begin{eqnarray}
F_1 & = & \overline{\Lambda} (\frac{A}{3} + v.p B) \nonumber \\
F_2 & = & \overline{\Lambda} (\frac{A}{3} +2v.p B) \label{rel2} \\
F_3 & = & \overline{\Lambda} B   \nonumber
\end{eqnarray}

We can now proceed to write down the semileptonic form-factors at
tree level and to order $1/m_Q$ in the heavy quark expansion. Using
(\ref{def}), (\ref{cur}), (\ref{mat1}), (\ref{mat2}) and
(\ref{rel2}) we have:

\begin{eqnarray}
f_+ + f_- & = & 2A\,m_{M}^{-1/2}[1-\frac{\overline{\Lambda}}{2m_M}
(1+4v.p\frac{B}{A})+\ldots] \label{sca3a}  \\
 & & \nonumber \\
f_+ - f_-& = & 2B\, m_{M}^{1/2}[1-\frac{\overline{\Lambda}}{2m_M}
+\ldots]   \label{sca3b}
\end{eqnarray}

It is interesting to analyze (\ref{sca3a}) and (\ref{sca3b}) in the chiral
limit, $v.p\rightarrow 0$. First, we observe that in this limit
the same scaling violation occurs in both the non-leading
(\ref{sca3a}) and the leading (\ref{sca3b}) combinations.
Moreover, the chiral couplings of heavy mesons to the
pseudo-goldstone bosons implies \cite{Wise,BD}

\begin{equation}
f_+ + f_- = \frac{f_M}{f_{\pi}}  \label{CT}
\end{equation}

We then conclude that in the chiral limit, the symmetry
breaking terms affecting (\ref{sca3a}) are the same as the ones
affecting the heavy meson decay constant $f_M$. Then, whatever the
size of the symmetry breaking efffects in $f_M$, equations
(\ref{sca3a}),
(\ref{sca3b}) and (\ref{CT}) tell us how they contaminate the
scaling procedure to extract $|V_{ub}|$ from $D\rightarrow \pi$ and
$B\rightarrow \pi$ data. The kinematic range shared by the two
processes corresponds to $v.p < 1$GeV, where (\ref{CT}) is
expected to be valid.

Large scaling violation effects in the heavy meson decay contant
are suggested by lattice
calculations as well as by QCD sum rules results. Thus, before
applying heavy quark symmetry arguments to phenomenology, these
calculations should be fully understood and under control.

Equations (\ref{sca3a}) and (\ref{sca3b}) were derived at tree
level. However the essential result, which relates the symmetry breaking
in the decay constant $f_M$ to the semileptonic form-factors, will
not change when going beyond tree level.

\section {Beyond Tree Level}

Beyond tree level there will be additional operators contributing
to the $1/m_Q$ expansion of the current. There will be an additional
dimension three operator, $Q_0'=\bar{q}v_{\mu}h_Q$. Also $Q_1$ will
mix with other dimension four operators. The vector current including
$1/m_Q$ corrections and $QCD$ corrections can be written as

\begin{equation}
\bar{q}\gamma_{\mu}Q= C_0(\mu)Q_0 + C_1(\mu)Q_0' +
\frac{1}{m_Q} \sum_{i=1}^{6} B_i(\mu) Q_i +\ldots   \label{cur2}
\end{equation}

\noindent
The $C_i$'s and $B_i$'s are the short distance coefficients of the
dimension three and dimension four operators respectively and they
introduce additional dependence on the heavy mass. Following
Ref.\cite{Neu}, we choose the basis for the dimension four operators
to be

\pagebreak
{\samepage
\begin{eqnarray}
Q_1 = &  \bar{q} \gamma_{\mu}i\!\!\not Dh_Q  & Q_4  =  (-iv.D)\bar{q}
\gamma_{\mu}h_Q \nonumber \\
Q_2  = & \bar{q}v_{\mu}i\!\not Dh_Q          & Q_5  =  (-iv.D)\bar{q}
v_{\mu}h_Q \label{bas} \\
Q_3  = & \bar{q} iD_{\mu}h_Q                 & Q_6  =  (iD_{\mu})
\bar{q}h_Q \nonumber
\end{eqnarray}}

The short distance coefficients are calculated in \cite{Mcor,Golden}. The
matrix elements of the dimension four operators in (\ref{cur2}) will
account for the subleading contributions coming from the expansion
of the current in the HQET.

Additional $1/m_Q$ corrections come from corrections to the heavy
quark propagator. These can be taken into account, when calculating
the matrix element (\ref{def}) in the HQET, as insertions
of the dimension four operators in the effective Lagrangian (\ref{eff3})
into the leading order current operator $Q_0$ \cite{Neu}

\begin{eqnarray}
  <\pi({\bf p})|i\int d^4xT[(\bar{q}\Gamma h_Q)_0,
(\bar{h}_Q(iD)^2h_Q)_x]|M({\bf v})> =  \nonumber \\
    G_1(\mu)
Tr [\gamma_5(A(\mu)\!\not v + B(\mu)\!\not p) \Gamma {\cal M}(v)] \label{ins1}
\end{eqnarray}

\begin{eqnarray}
<\pi({\bf p})|i\int d^4xT[(\bar{q}\Gamma h_Q)_0,
(\bar{h}_Q\sigma^{\mu \nu}G_{\mu \nu}h_Q)_x]|M({\bf v})> =  \nonumber \\
   Tr [W^{\mu \nu}\gamma_5(A(\mu)\!\not v + B(\mu)\!\not p) \Gamma
(\frac{1+\not v}{2})\sigma_{\mu \nu}{\cal M}(v)]  \label{ins2}
\end{eqnarray}

\noindent
where $T$ denotes time-ordered product and $W^{\mu \nu}$ is an
antisymmetric tensor that can be written as

\begin{equation}
W^{\mu \nu}=G_2(\mu)\sigma^{\mu \nu} + G_3(\mu)(\gamma^{\mu}p^{\nu}-
\gamma^{\nu}p^{\mu})  \label{Wmn}
\end{equation}

\noindent
$G_1(\mu)$ and $G_2(\mu)$ are new long-distance parameters
evaluated at the energy scale $\mu$. The matrix elements of $Q_2$
and $Q_3$ can be written as in (\ref{mat2}) and then their
computation is similar to $Q_1$ with different Dirac structure. The
matrix elements of the
operators $Q_4$, $Q_5$ and $Q_6$ are related to those by using
(\ref{parts}).
The result, to all orders in $\alpha_s$ and to order $1/m_Q$
in the HQET is

\begin{eqnarray}
f_+ + f_- & = & 2A(\mu)m_{M}^{-1/2}(C_0(\mu)+C_1(\mu))
[1-\frac{\overline{\Lambda}}
{2m_M}B_+(\mu) + \frac{G_+(\mu)}{m_M}]  \label{sca4a} \\
 & & \nonumber \\
f_+ - f_- & = & 2B(\mu)m_{M}^{1/2}C_0(\mu)[1-\frac{\overline{\Lambda}}
{2m_M}B_-(\mu) + \frac{G_-(\mu)}{m_M}]  \label{sca4b}
\end{eqnarray}

\noindent
Here we defined

\begin{eqnarray}
B_+ & = & B_1(1+4v.p\frac{B}{A})+B_2(1+v.p\frac{B}{A})+B_3v.p\frac{B}{A}
\nonumber \\
    &   & +2B_4(1+2v.p\frac{B}{A})+B_5(1+\frac{5}{2}v.p\frac{B}{A})
+B_6(1+v.p\frac{B}{A}) \label{B+} \\
B_- & = & B_1+B_3+2B_4+B_6 \label{B-} \\
G_+ & = & G_1+16G_2+4G_3m_{\pi}^{2} \label{G+} \\
G_- & = & G_1+16G_2-8G_3v.p  \label{G-}
\end{eqnarray}

\noindent
The situation looks, in principle, more complicated than at tree level.
When taking the chiral limit to make a connection with the heavy meson
decay constant via ({\ref{CT}), we observe that the contributions from
the $1/m_Q$ corrections to the hadronic wave function -$G_1(\mu)$
and $G_2(\mu)$-
as well as the contributions from $Q_1$, $Q_4$ and $Q_6$ are the same
in (\ref{sca4a}) and (\ref{sca4b}). However the symmetry breaking
induced by $Q_2$, $Q_3$ and $Q_5$ differs. Also the short distance
coefficient $C_1(\mu)$, corresponding to $Q_0'$, only affects
$f_++f_-$.

\noindent
Nevertheless, the essence of the result from the previous section
is still present: in the chiral limit the symmetry breaking terms
in (\ref{sca4a})are the same as the ones in the heavy meson
decay constant $f_M$. The scaling violating short and long distance
parameters entering in (\ref{sca4a}) and (\ref{sca4b}) are the same
(and, to a very good approximation, enter in the same way in both).
Then, any calculation addressing the scaling violations in $f_M$
is at the same time a calculation of the effect in the
semileptonic form-factors. For
instance, to make contact with the work on $f_M$ in Ref.\cite{Neu}
we can define $A(\mu)=F(\mu)/2f_{\pi}$, with $F(\mu)$ defined in
\cite{Neu}. Thus the QCD sum rules results for $F(\mu)$, $G_1(\mu)$
and $G_2(\mu)$ directly apply to the semileptonic
form-factors in (\ref{sca4a}) and (\ref{sca4b}).

\section {Conclusions}
We have calculated the semileptonic form-factors entering the
decay $H\rightarrow \pi e \nu$ to order $1/m_Q$ in the HQET,
both at tree level and to all orders in QCD. By using
equation~(\ref{CT}), a result of the unified heavy quark and
chiral symmetric formulation, we have shown that the same symmetry
breaking terms affecting the heavy meson decay constant $f_M$
enter in both the non-leading and the leading combinations of
semileptonic form-factors.

Equations (\ref{sca4a}) and (\ref{sca4b}) help identify
the origin of the most important scaling violations. The
long~distance parameters should be computed on the lattice
or by $QCD$ sum rules, as applied to $f_M$. Only when they
are known with confidence in these or any other calculational
scheme, will the extraction of $|V_{ub}|$ from $D,B\rightarrow
\pi e \nu$ plus mass scaling be a phenomenologically sensible
procedure.

The extension of the calculation to include final states with
light vector mesons ($\rho$,~$K^*$,~etc) is straightforward.
Here, however, there does not seem to be an equivalent to
equation~(\ref{CT}) relating $f_M$ with the semileptonic
form-factors. The $1/m_Q$ corrections arising from the
operators in  (\ref{bas}) are still the same (they only depend
on the long distance parameter $\overline{\Lambda}$);
but the ones
induced by insertions like those in (\ref{ins1},\ref{ins2}) need
not be related, in principle, to the equivalent symmetry
breaking terms in $f_M$.
The formulation of the chiral couplings of heavy mesons to
light vector mesons \cite{Casal} does not seem to provide, by
itself, such a connection.

\vspace{0.2in}
{\bf Acknowledgements:} The author thanks John F. Donoghue for
useful discussions and suggestions. This work was supported in part by
the U.S. National Science Foundation.

\end{document}